\documentstyle[epsf,prl,aps,preprint,tighten]{revtex}
\newcommand{\beq}{\begin{equation}}
\newcommand{\eeq}{\end{equation}}
\newcommand{\bea}{\begin{eqnarray}}
\newcommand{\eea}{\end{eqnarray}}

\def\a{\alpha}
\def\b{\beta}

\def\ch{\chi}

\def\ph{\phi}
\def\s{\sigma}
\def\th{\theta}

\newcommand{\no}{\nonumber \\}
\def\th{\theta}

\preprint{\hfill\parbox{4cm}{UT-KOMABA 00-13\\}}

\begin{document}
\draft \title{Tunneling between the giant gravitons in $AdS_5 \times S^5$ }
\author{Julian Lee\footnote{jul@hep1.c.u-tokyo.ac.jp}}
\address{Institute of Physics, University of Tokyo \\                
 Komaba, Meguro-ku, Tokyo 153, Japan}
\date{\today}
\maketitle 
\begin{abstract}
 I consider the giant gravitons in $AdS_5 \times S^5$. By numerical simulation, I show a strong indication that there is no instanton solution  describing the direct tunneling between the giant graviton in the $S^5$ and its dual counterpart in the $AdS_5$. I argue that it supports the supersymmetry breaking scenario suggested in ref.\cite{mye} 
\end{abstract}

\setcounter{footnote}{0}
\narrowtext 

%\newpage
%\twocolumn
%\vspace{0.5cm}
 Stable extended brane configurations in  spaces of the type $AdS_m \times S^n$, called giant gravitons\cite{mye,sus,has,je1,je2}, are interesting  in connection with the $AdS/CFT$ correspondences\cite{mal}. It was suggested\cite{sus} that they might provide a  mechanism to realize the so-called ``stringy exclusion principle"\cite{string}. According to this principle, which is shown in the conformal field theory side, there is an upper bound to the weight  of a chiral primary state. For the duality between the supergravity on $AdS$ space and the boundary conformal field theory to hold, we should be able to see this in the gravity side too, weight being mapped to the angular momentum in $S^n$. In fact, a spherical D or M(n-2)-brane  embedded in the  $S^n$ part of $AdS_m \times S^n$ was considered in ref.\cite{sus}, which has the same quantum numbers as a point-like Kaluza-Klein excitation. It was shown that the size of this configuration grows with the angular momentum. Since the angular  momentum is now bounded by the radius of $S^n$, this seemed to provide a natural explanation for the stringy exclusion principle.  However, a puzzle remained since there are also a point-like state, and a stable extended brane embedded in $AdS_m$ space\cite{mye,has}, called `(dual) giant graviton', which are degenerate with the giant graviton in $S^n$. Obviously, they do not have any upper bound on the angular momentum. One possible way out of this puzzle was suggested in ref.\cite{mye}. It was suggested that due to the quantum tunneling, the degeneracy of these  states might get lifted, and there might be no supersymmetric state when the angular momentum exceeds the upper bound. In this light it is interesting to study instanton solutions connecting these configurations, which might break supersymmetry via the tunneling effect.  In fact, the instantons describing the tunneling between a giant graviton  and the point-like graviton were found in analytic form for both $AdS_m$ and $S^n$ cases\cite{mye,has,je1}. The values of the Euclidean action for these solutions were obtained, and their supersymmetry(SUSY) properties were investigated. It was found that they are $\frac{1}{4}$ BPS states\cite{has}, whereas the giant gravitons and the point-like one are all $\frac{1}{2}$ BPS states.  

 It would be interesting to see  whether there is  also an instanton solution describing the direct tunneling between the two types of giant gravitons. We consider this problem  for the case of $AdS_5 \times S^5$. This is the simplest case to work on with the test brane formalism, since the dimensionalities of the initial and the final branes are the same.  By numerical simulation, we show a strong indication that there is no instanton solution directly connecting the two giant gravitons directly. We  argue that this supports the SUSY breaking scenario suggested in ref.\cite{mye}.

To begin with, we consider a D3-brane living in $AdS_5 \times S^5$ space, whose metric is of the form,

\beq
ds^2  = ds_{AdS}^2 + ds_{S}^2
\eeq
with
\beq
ds_{AdS}^2 = -\cosh^2 \beta dt^2 + L^2 d\beta^2 +L^2 \sinh^2 \beta (d\a_1^2 + \sin^2 \a_1  ( d\a_2^2 + \sin \a_2^2 d\a_3^2)) \label{adsm}
\eeq
and
\beq
ds_{S}^2 = L^2\cos^2 \theta d\ph^2 + L^2 d\th^2 +L^2 \sin^2 \th (d\ch_1^2 + \sin^2 \ch_1 ( d\ch_2^2 + \sin \ch_2^2 d\ch_3^2)), \label{sm}
\eeq
$L$ being the scale of the $AdS_5$ and $S^5$.
 The background gauge field components in the orthonormal frame is given by
\beq
F_{\hat t \hat r \hat \a_1 \hat \a_2 \hat \a_3} = F_{\hat \ph \hat \th \hat \ch_1 \hat \ch_2 \hat \ch_3}= -\frac{4}{L}.
\eeq  
and the corresponding gauge fields in the coordinate basis are
\bea
A_{t \a_1 \a_2 a_3} &=& L^3 \sinh^4 \b \sin^2 \a_1 \sin \a_2 \no
A_{\ph \ch_1  \ch_2 \ch_3} &=& L^3 \sin^4 \th \sin^2 \ch_1 \sin \ch_2. 
\eea

We now consider the worldvolume action of the D3-brane embedded in this background, which is given by:
\beq
S = T\int d\tau \prod_{i=1}^3 d\sigma_i ( - \sqrt{-g} +  A_{M_1 M_2 M_3,M_4}\frac{\partial X^{M_1}}{\partial \tau}\frac{\partial X^{M_2}}{\partial \s_1} \frac{\partial X^{M_3}}{\partial \s_2}\frac{\partial X^{M_4}}{\partial \s_3})
\eeq 
where $g$ is the determinant of the induced metric on the world-volume,
\beq
g_{i j } = \frac{\partial X^M}{\partial x^i}\frac{\partial X^N}{\partial x^j} G_{M N} 
\eeq
with $G_{M N}$ being the background metric, $X^M$ the coordinates on $AdS_5 \times S^5$, ($\tau,\sigma^i$) the coordinates on the world-volume,   and $T$ the tension of the brane. We set the world-volume gauge field and the fermions to zero, which is a consistent ansatz when we consider a classical solution, since they appear in the action from the quadratic order. 

\begin{figure}
\epsfxsize=9.00cm
\epsfysize=5.00cm
\vskip0.1cm
\centerline{
\epsfbox{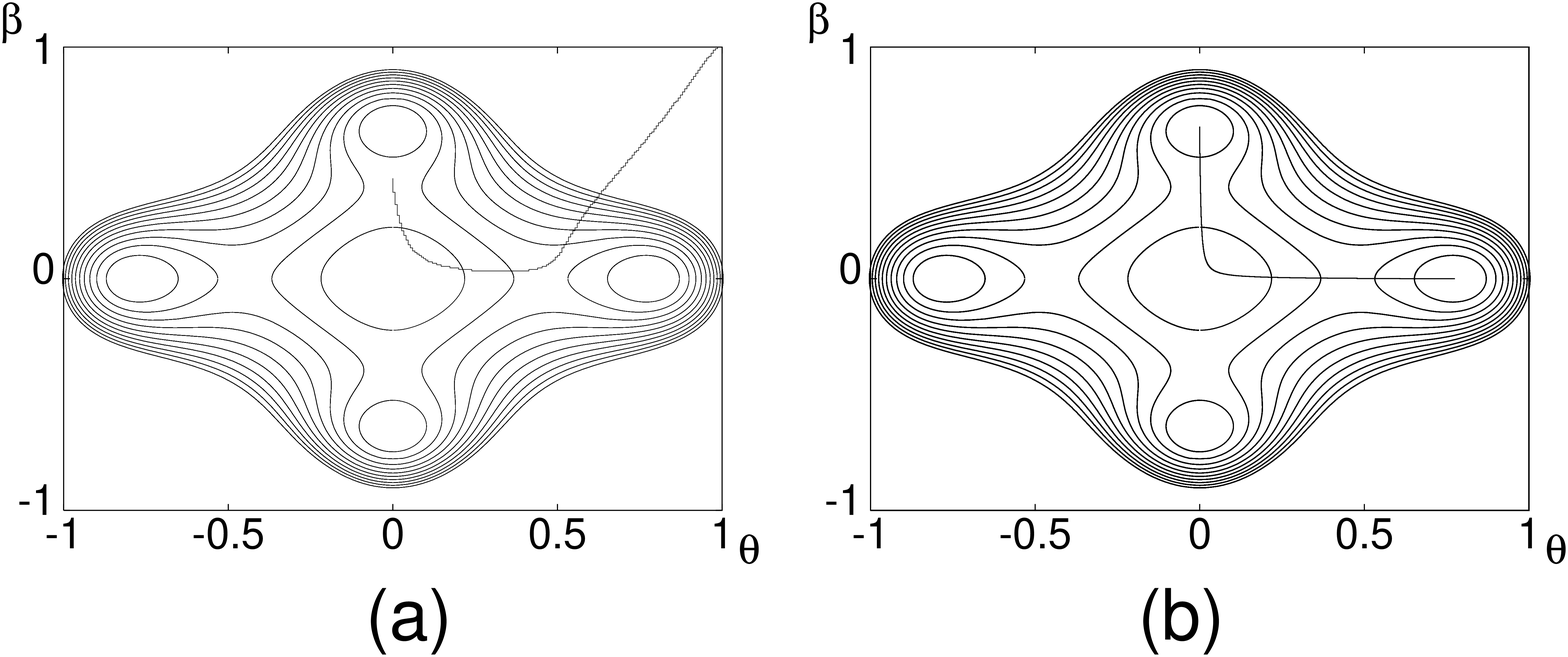}}
\vskip0.1cm
\caption{ (a) The equipotential contours and the particle trajectory for $m=0.5, \delta \b=0.0001$. The contours are drawn for the potential differences $\Delta V =0.0005$. The regions $\th<0$ or $\b<0$ are redundant, but drawn for the sake of clarity. Unless $v_\th/v_\b \ll 1$ initially, the particle bounces to infinity before reaching the other potential hill. (b) The particle reaches the final hill-top only if it passes near by the origin. }
\label{ocon}  
\end{figure}

 We consider a configuration where the coordinates of the D3-brane have components both in $AdS_5$ and $S^5$, and projections to $AdS_5$ and $S^5$ both have the spherical topology. Then it is convenient to use the gauge where we set
\bea
\a_i &=& \ch_i = \sigma_i \no
t &=& \tau.
\eea 
 We can also consistently set $\frac{\partial \th}{\partial  \sigma_i}$,$\frac{\partial \b}{\partial  \sigma_i}$, and $\frac{\partial \ph}{\partial  \sigma_i}$ to zero, since they are at least of quadratic order. After angular integrations, we get the Lagrangian:
\beq
{\cal L} =- \frac{N}{L}\sqrt{\cosh^2 \b -L^2  \dot \b^2 -L^2  \dot \th^2 - L^2 \dot \ph^2 \cos^2 \theta } (\sinh^2 \b + \sin^2 \th)^{3/2} + \frac{N}{L} \sinh^4 \b  + N \dot \ph \sin^4 \th .
\eeq
where we set $T=\frac{N}{2\pi^2 L}$ using background flux quantization condition. $N$ is an integer corresponding to the number of D3-branes which are the sources of the background spacetime.  We obtain the canonical momenta in usual way,
\bea
P_\theta &\equiv& \frac{\partial \cal L}{\partial \dot \theta}= \frac{NL \dot \theta (\sinh^2 \b + \sin^2 \th)^{3/2}  }{ \sqrt{\cosh^2 \b - L^2 \dot \b^2 - L^2 \dot \th^2 - L^2  \dot  \ph^2 \cos^2 \theta } } \no
P_\b &\equiv& \frac{\partial \cal L}{\partial \dot \b}= \frac{NL \dot \b (\sinh^2 \b + \sin^2 \th)^{3/2}  }{ \sqrt{\cosh^2 \b - L^2 \dot \b^2 - L^2 \dot \th^2 - L^2 \dot \ph^2\cos^2 \theta } } \no
P_\ph &\equiv& \frac{\partial \cal L}{\partial \dot \ph}= \frac{NL \dot \ph \cos^2 \th (\sinh^2 \b + \sin^2 \th)^{3/2} }{ \sqrt{\cosh^2 \b - L^2 \dot \b^2 - L^2 \dot \th^2 -L^2  \dot \ph^2 \cos^2 \theta } } +N  \sin^4 \theta  .
\eea
We see that since the Lagrangian has no $\ph$ dependence, $P_\ph$ is a conserved quantity.  The Hamiltonian is obtained by the Legendre transformation,
\beq
H =  -\frac{N}{L}\cosh \b \sqrt{ (\sinh^2 \b + \sin^2 \th)^3  - p_\b^2 - p_\th^2 + \frac{(p -\sin^4 \th)^2}{\cos^2 \th}} +\frac{N}{L}\sinh^4 \b . \label{ham}
\eeq
where $p \equiv \frac {P}{N}$ for all the coordinates. We also made the Euclidean continuation $p_\b^2$, $p_\th^2 \to -p_\b^2, -p_\th^2$ and $H \to -H$.  The problem is now reduced to a two dimensional relativistic classical mechanics, $p_\phi$ being treated as a c-number which we write as $m$ from now on.  The instanton solution is found by solving the Hamiltonian equation
\beq
\dot q^i = \frac{\partial H}{\partial p_i} \qquad \dot p_i = -\frac{\partial H}{\partial q^i} \label{heq}
\eeq
with $q^i=(\b,\th), p_i=(p_\b, p_\th)$. 
We could not  solve this equation analytically, so we made a numerical simulation. In fact, since (\ref{heq})  is a first-order differential equation, there is always unique solution to a given initial condition.   By energy conservation, the particle with energy $E$ moves only in the region  $V(\th,\b)<E$, where the potential is defined as
\beq
V(\th,\b) \equiv H(p_\th=p_\b=0).
\eeq
Therefore the contour $V=E$  is the set of turning points. The equal-height contour of the potential is depicted for  $m=0.5$ in Fig.\ref{ocon}. The general features are the same for other values of $m$. In particular, there are three potential hills  corresponding to the zero-sized brane at the origin, and the two kinds of giant gravitons at $(\th, \b)=(\sin^{-1} \sqrt{m},0)$ and $(\th, \b)=(0,\sinh^{-1} \sqrt{m})$. An instanton solutions should have the boundary conditions such that it is at one of the hill-tops at $t \to -\infty$, and reaches another hill-top at $t\to \infty$.  In order to impose the boundary condition at $t\to -\infty$, we note that in this limit the problem reduces to that of the non-relativistic mechanics in the presence of an inverted harmonic potential. Assuming the initial configuration at $t\to -\infty$ is a (dual) giant graviton, we have 
\bea
\frac{p_\th}{m} \simeq \dot  \th &\simeq & \ 2 \delta \th \no
\frac{p_\b}{m} \simeq \dot  \b &\simeq &  2  \delta \b \label{lin}
\eea
where the $\delta \th$ and $\delta \b$  are the perturbations around the potential hill.  Here we used the fact
\bea
\frac{\partial^2 V}{\partial \th^2}&=&\frac{\partial^2 V}{\partial \b^2}=2m . \no
\frac{\partial^2 V}{\partial \th \partial \b}&=& 0.
\eea
for the potential hills corresponding to the giant gravitons. We then tune the direction of the initial velocity, $j \equiv \delta \dot \th/  \delta \dot \b= \delta \th/\delta \b$ so that the particle reaches another hill-top at $t \to \infty$.  The evolution was done using the 4-th order Runge-Kutta methods, where we approximate
\beq
f(t+\Delta t) \simeq  \sum_{n=0}^4 \frac{(\Delta t)^n}{n!}\frac{\partial^n}{\partial t^n}f(t)  
\eeq
with time interval $\Delta t=0.0001L$, in the Java programming language.
We could reproduce the instanton solution connecting one of the giant gravitons and the point-like solution\cite{mye,has,je1} by shooting the particle directly toward the origin. We could see that the graph was in almost perfect agreement with the known analytic form, for the motion between the hill tops.  We also checked the energy conservation for arbitrary trajectories, and found that the energy is indeed constant without any visible deviation. These results assure us of the accuracy of our simulations.  We then use this simulation method to look for the instanton solution connecting two giant gravitons directly, by changing the direction of the initial velocity.   Typical behavior of our particle for several initial directions are depicted in Fig.1, for $m=0.5$. The qualitative behaviors are the same for other values of $m$. 

Surprisingly, we  find that in our particle reach the other potential hill only if we shoot the particle almost toward the origin. That is, the solution looks like the sum of two instantons connecting the giant gravitons with the point-like configuration. In fact, as we  decrease $\delta \th, \delta \b$ to increase the time spent near the initial hill top, we have to tune the direction of the initial velocity more and more toward the origin in order for the particle to reach the final hill-top, which makes particle spend longer time near the origin. Therefore  we see that our solution is just a remnant of the two-instanton effect. 

Since this is the only way one can make our particle reach the final hill-top, we conclude that there is no instanton solution describing the direct tunneling between the giant gravitons.

To strengthen our statement, we consider a special limit, $m<< 1$. Since the coordinates of the giant gravitons are of order $O(\sqrt{m})$, and we are interested in the motion of the particle  between them, we have
\bea
\th \sim \b &\sim& O(\sqrt{m})  \no
p_\th \sim p_\b &\sim& O(m^{3/2}), 
\eea
\begin{figure}
\epsfxsize=9.00cm
\epsfysize=5.00cm
\vskip0.1cm
\centerline{
\epsfbox{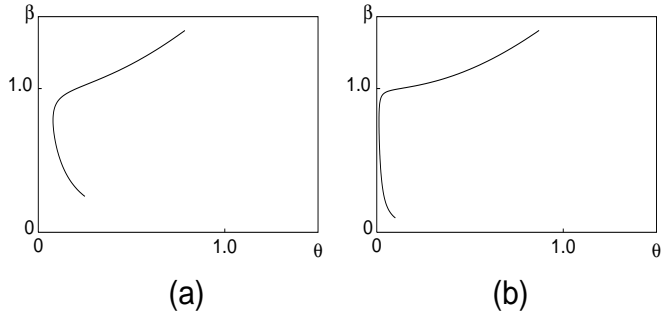}}
\vskip0.1cm
\caption{ (a) The particle trajectory in the non-relativistic limit $m << 1$, for $r_0= 0.25$. $m$ is now scaled out from the problem.  $\b < \th$ portion is just the reflection of this picture.  (b) Similar picture for $r_0= 0.1$. We see that the particle reaches the hill-top only in the limit $r_0 \to 0$.  }
\label{nr}  
\end{figure}
and to the leading order in $m$, the Hamiltonian Eq.(\ref{ham}) reduces to
\beq
H = \frac{p_\th^2+p_\b^2}{2m}-m(\frac{\th^2+\b^2}{2})+\th^4+\b^4-\frac{(\th^2+\b^2)^3}{2 m}
\eeq
which is now a non-relativistic system. We subtracted the constant $m$. The Hamiltonian above is of order $O(m^2)$, and we have thrown away the terms of $O(m^3)$ and higher. We then solve the Hamiltonian equation (\ref{heq}) with this Hamiltonian. We see that the $m$ can be scaled out by redefining 
\beq
\th, \b \to \sqrt{m}\th, \sqrt{m} \b.
\eeq 
The potential hills are now at $(\th, \b)=(0,0), (1,0), (0,1)$. Because of the symmetry under the exchange the coordinates $\th \leftrightarrow \b$, we see that the trajectory of an instanton solution between the $S^5$ and $AdS_5$ giant gravitons should intersect the line $\th = \beta$ orthogonally. Also, the magnitude of the velocity $v_0$ at this point is given by the energy conservation,
\beq
\frac{v_0^2}{2}-\frac{r_0^2}{2}+\frac{r_0^4}{2}-\frac{r_0^6}{2}=0
\eeq
where $r_0$ is the distance of this point from the origin. Therefore we only have to change the parameter $r_0$ and see whether our particle reaches the hill-top. We note that it reaches the hill-top only in the limit $r_0 \to 0$. Therefore again, we see that the only solution connecting the giant gravitons is the two-instanton solution. This result is plotted in Fig.\ref{nr} for several values of $r_0$.
   
These results indicate that the tunneling between the giant gravitons happen only through the instantons connecting them with the point-like configuration. This implies that the system is effectively one-dimensional, as far as instanton effects are concerned. In particular, the SUSY breaking scenario suggested in ref.\cite{mye} can now be made more concrete, by making analogy with the case of one-dimensional supersymmetric quantum mechanics with double and triple potential well\cite{w,sq,by,inst}. The  crucial ingredient in SUSY breaking and restoration in the one-dimensional quantum mechanics is the fermionic zero modes\cite{fz}, which make neighboring instanton and (anti)-instanton form so-called instanton molecule\cite{by}.  

  This is because whenever an (anti)-instanton is separated very far from the rest, the fermionic zero mode around this (anti)-instanton makes the path-integral vanish.  We expect that this qualitative picture does not change for our system, the only difference being the number of fermionic zero modes. Then the calculation would be almost identical to the one-dimensional quantum mechanics. As we sum the multi-instanton configuration, any contribution from the odd number of (anti-) instantons  vanish. Also, the fact that (anti-) instantons move in pairs gives different combinatorics from the case of the dilute instanton gas. When $m > 1$, we expect that the Hamiltonian matrix between the two minima would be\cite{inst}
\beq 
\Delta H=\pmatrix{K e^{-2A} I&0\cr 
                0& Ke^{-2A} I\cr}
\eeq
where $A$ is the Euclidean action of the instanton between the $AdS_5$ giant graviton and the zero-sized brane, K is the  determinant due to the  non-zero fluctuation modes, $I$ and $0$ are the $2^8$ dimensional identity and zero matrices respectively.  We see that the energy is lifted and consequently the SUSY is broken. When $m<0$, the instanton molecule picture implies that the Hamiltonian matrix between the three minima is given by
\beq
\Delta H=\pmatrix{K e^{-2A}I&0& \sqrt{K \tilde K}e^{-A-B}I\cr 
                0&(K e^{-2A}+\tilde K e^{-2B})I&0\cr
                \sqrt{K \tilde K} e^{-A-B} I&0&\tilde K e^{-2B}I}
\eeq
 where now $B$ is the Euclidean action of the instanton between the $S^5$ giant graviton and the zero-sized brane,
 and $\tilde K$ is again the determinant due to the non-zero fluctuation modes around it. We now see that there is  a short multiplet whose  energy is not lifted and therefore still supersymmetric. 
   
Note that it was very crucial in this argument that there is no direct tunneling between the giant graviton in $S^5$ and its dual counterpart in $AdS_5$. Had there been such an instanton, the picture of SUSY breaking would have been very obscure.

Of course it remains to be explicitly shown that the instantons really form molecules due to the fermionic zero modes. Also, we need to prove that there is no instanton solution which either is not spherical or has non-vanishing worldvolume gauge fields. These issues are left for the future investigations.  

The numerical simulations in this paper can be performed at  http://hep1.c.u-tokyo.ac.jp/\~{}jul/giant1/Applet1.html .

\acknowledgments

The author thanks Antal Jevicki, Tamiaki Yoneya, Seungjoon Hyun, Sumit Das, \O yvind Tafjord, and Piljin Yi  for useful discussions and comments.  
This work was supported  by the Japanese Society for Promotion of Science through the Institute of Physics, University of Tokyo.


\begin{thebibliography}{99}
\bibitem{mye} M. T. Grisaru, R. C. Myers and O. Tafjord, JHEP {\bf 0008} (2000) 040, hep-th/0008015.
\bibitem{sus} J. McGreevy, L. Susskind and N. Toumbas, JHEP {\bf 0006} (2000) 008, hep-th/0003075.
\bibitem{has} A. Hashimoto, S. Hirano and N. Itzhaki, JHEP {\bf 0008} (2000) 051, hep-th/0008016.
\bibitem{je1} S. Das, A. Jevicki and S. D. Mathur, hep-th/0008088.
\bibitem{je2} S. Das, A. Jevicki and S. D. Mathur, hep-th/0009019. 
\bibitem{mal} J. Maldacena, Adv. Theor. Math. Phys. {\bf 2} (1998) 231, hep-th/9711200.
\bibitem{string} J. Maldacena and A. Strominger, JHEP {\bf 9812} (1998) 005, hep-th/9804085;  A. Jevicki and S. Ramgoolam, JHEP {\bf 9904} (1999) 032, hep-th/9902059; P. Ho, S. Ramgoolam and R. Tatar, Nucl. Phys. {\bf B573} (2000) 364, hep-th/9907145;
S. S. Gubser, Phys. Rev. {\bf D56} (1997) 4984, hep-th/9704195. 
\bibitem{po} H. Lu, C. N. Pope and J. Rahmfield, J.Math.Phys. {\bf 40} (1999) 4518-4526, hep-th/9805151.
\bibitem{fz} H. Bohr, E. Katznelson and K. S. Narain, Nucl. Phys. {\bf B238} (1984) 407; J. M. Leinaas, K. Olaussen, Nucl. Phys. {\bf B239} (1984) 209.
\bibitem{w} E. Witten, Nucl. Phys. {\bf B188} (1981) 513; 
\bibitem{sq} P. Salomonson and J. W. Van Holten, Nucl. Phys. {\bf B196} (1982) 509; D. Lancaster, Nuovo Cim. {\bf A79} (1984) 28;  F. Cooper, A. Khare, and U. Sukhatme,
Phys.Rept. {\bf 251} (1995) 267-385.
\bibitem{by} I. I. Balitsky and A. V. Yung, Nucl. Phys. {\bf B274} (1986) 475. 
\bibitem{inst} R. K. Kaul and L. Mizrachi, J. Phys. {\bf A 22} (1989) 675. 

\end{thebibliography}
\end{document}